\documentclass[epj]{svjour}

\usepackage{bm,graphicx,stmaryrd,color}

\newcommand{\hi}{h_i}
\newcommand{\hij}{h_{ij}}
\newcommand{\intomeg}{\int_{\Omega}\!d^2r}
\newcommand{\intdeltaomeg}{\int_{\delta\Omega}\!d^2r}
\newcommand{\intderonomeg}{\int_{\mathcal{B}}\!}
\newcommand{\Oe}{\mathcal{O}\!\p{\epsilon}}
\newcommand{\OO}{\mathcal{O}\!\p{\epsilon^2}}
\newcommand{\OOO}{\mathcal{O}\!\p{\epsilon^3}}
\newcommand{\OOOOp}{\mathcal{O}'\!\!\p{\epsilon^4}}

\newcommand{\ghc}{\p{\bm{\nabla}h}^2}
\newcommand{\egg}{\mathrel{\rlap{\raise1.7pt\hbox{$\,^2$}}{\hbox{$=$}}}}

\newcommand{\p}[1]{\left({#1}\right)}
\newcommand{\psq}[1]{\left[{#1}\right]}

\newcommand{\deron}[2]{\frac{\partial{#1}}{\partial{#2}}}
\newcommand{\fder}[2]{\frac{\delta{#1}}{\delta{#2}}}

\begin{document}
\title{Membrane stress tensor in the presence of lipid density and composition inhomogeneities}

\author{Anne-Florence Bitbol\inst{1} \and Luca Peliti\inst{2} \and Jean-Baptiste Fournier \inst{1}\fnmsep\thanks{Author for correspondance.}}
\institute{Laboratoire Mati\`ere et Syst\`emes Complexes (MSC),
Universit\'e Paris Diderot---Paris 7 \& UMR CNRS 7057, 10 rue Alice Domon et L\'eonie Duquet, F-75205 Paris Cedex 13, France
\and
Dipartimento di Scienze Fisiche
and Sezione INFN, Universit\`a ``Federico II",
Complesso Monte S. Angelo, I-80126 Napoli, Italy
}

\abstract{We derive the expression of the stress tensor for one and two-component lipid membranes with density and composition inhomogeneities. We first express the membrane stress tensor as a function of the free-energy density by means of the principle of virtual work. We then apply this general result to a monolayer model which is shown to be a local version of the area-difference elasticity (ADE) model. The resulting stress tensor expression generalizes the one associated with the Helfrich model, and can be specialized to obtain the one associated with the ADE model. Our stress tensor directly gives the force exchanged through a boundary in a monolayer with density and composition inhomogeneities. Besides, it yields the force density, which is also directly obtained in covariant formalism. We apply our results to study the forces induced in a membrane by a local perturbation.
}

\date{\today}
\maketitle

%%%
\section{Introduction}
\label{intro}

Biological membranes consist principally of a lipid bilayer.
In each monolayer, the molecules are arranged with their hydrophobic ends directed inward toward the second monolayer, and their hydrophilic ends directed outward toward the surrounding water~\cite{Alberts_book,Mouritsen_book}. Membranes are fluid and possess a weak resistance to bending, so they fluctuate strongly under the effect of thermal excitations~\cite{Helfrich73}. In many relevant situations, their total area may be treated as a constant: this is due to the existence of a practically constant area per lipid and to the fact that lipids are virtually insoluble in water~\cite{Mouritsen_book}. As a consequence, membranes develop a lateral tension in response to external forces~\cite{Helfrich84,Fournier01}.

Although the constant area approximation works finely for many fundamental aspects (see, e.g., Refs.~\cite{Helfrich84,Helfrich78,Brochard75,Engelhardt85,Meleard92,Lim02,Derenyi02,Baumgart03}), there are important situations in which it is essential to take into account the variability of the lipid density
{\textit{in each monolayer}. Indeed,
an essential parameter in the phase diagram of vesicle shapes is the difference between the areas of the two monolayers~\cite{Evans80,Svetina89,Wiese92,Miao94}, and the most accurate equilibrium description of the elasticity of membranes thus involves a monolayer ``area difference elasticity" (ADE) term in addition to the usual bending terms~\cite{Svetina85,Miao94,Seifert_book}. There are also
spectacular instabilities that may be triggered by a local expansion of one monolayer, yielding cristae-like invagination~\cite{Khalifat08}, ejection of tubules~\cite{Fournier09} or bursting and curling of polymersomes~\cite{Mabrouk09}. Finally, the dynamics of membranes at the sub-micron scale is controlled by the intermonolayer friction resulting from local velocity differences between the monolayers fluids that must be treated as compressible~\cite{Evans94,Seifert93}.

In this paper, we derive a fundamental tool that will help to rationalize these phenomena: the stress tensor acting in a membrane \textit{monolayer} of variable shape, variable lipid density and variable composition. 
Understanding forces in complex membranes with various degrees of freedom is crucial to understand their equilibrium shape and their dynamics \cite{Kabaso10,Napoli10}. 
The divergence of the stress tensor gives the density of elastic forces in the membrane, which is the basis of a dynamical description. It is very useful in the study of shape instabilities \cite{FuturMig}. Moreover, the stress tensor provides the forces directly exchanged through a boundary. This is a valuable information that can be used, e.g., in the calculation of membrane-mediated interactions~\cite{Muller07,Bitbol10}, or in the study of the spatial distribution of forces on the edge of an object embedded in the membrane.

The outline of our paper is the following: In Sec.~2, we present a formal derivation of our stress tensor from the principle of virtual work. In Sec.~3, following Ref.~\cite{Seifert93}, we construct a free-energy density for one and two-component monolayers that extends that of the ADE model. In Sec.~4, we derive the stress tensor associated with this free-energy density, and in Sec.~5 we discuss the associated force density. Finally, in Sec.~6, we show how our results enable to understand the forces and the dynamics in the case of a local perturbation of a membrane by the microinjection of a reagent.

\section{Stress tensor formal derivation}
\label{formal}

Let us consider one monolayer of the bilayer. We describe its shape in cartesian coordinates by the equation $z=h(x,y)$ of a surface $\mathcal{S}$ parallel to its hydrophobic interface with the other monolayer. Such a description, based on the height of the membrane with respect to a reference plane, is often referred to as the Monge gauge. We do not assume yet that the membrane is weakly deformed. Let $\bar\rho(x,y)$ be the \textit{projected} mass density, i.e., the lipid mass per unit area of the reference plane $(x,y)$. In order to study the case of a two-component monolayer, let us denote by $\phi(x,y)$ the local mass fraction of one of the two lipid species, say species number 1. The case of a one-component monolayer can be obtained by setting $\phi=0$.

Let $\bar f(\bar\rho,\phi,\hi,\hij)$ be the projected free-energy density of the monolayer (i.e., the free energy per unit area of the reference plane). Here and in the following, Latin indices represent either $x$ or $y$ (not $z$) and $\hi\equiv\partial_ih$, $\hij\equiv\partial_i\partial_jh$, etc. Note that we are assuming that the free energy depends only on the mass density, on the local lipid composition, and on the slope and curvature of the monolayer. We have thus neglected the gradients of the curvature and of the lipid density and composition. The former approximation has already proved successful~\cite{Helfrich73,Miao94}; the latter is justified by the fact that the correlation length of the density fluctuations should not be larger than the monolayer thickness (far from a critical point). 

Let us consider an infinitesimal cut with length $ds$ separating a region $\mathcal{A}$ from a region $\mathcal{B}$ in the monolayer, and let us denote by $\bm{m}$ the normal to the \textit{projected} cut directed toward region~$\mathcal{A}$. The \textit{projected stress tensor} $\bm{\Sigma}$ relates linearly the force $d\bm{f}$ that region $\mathcal{A}$ exerts onto region $\mathcal{B}$ to the vectorial length $\bm{m}\,ds$ of the projected cut through
\begin{equation}
d\bm{f}=\bm{\Sigma} \cdot \bm{m} \, ds\,.
\end{equation}
This defines the six components of the projected stress tensor: $\Sigma_{ij}$ and $\Sigma_{zj}$, where $i\in\{x,y\}$ and  $j\in\{x,y\}$~\cite{Fournier07}.

To determine the projected stress tensor, we shall follow the method presented in Ref.~\cite{Fournier07}, which is based on the principle of virtual work.
Let us consider a monolayer patch standing above a domain $\Omega$ of the reference plane. This patch is supposed to be a \emph{closed system} with fixed total mass of each lipid species. Its free energy reads 
\begin{equation}
 F=\intomeg\,\bar f\p{\bar\rho,\phi,\hi,\hij}\,.
\end{equation}

In order to deal with arbitrarily deformed states of the monolayer patch, we
assume that in addition to the boundary forces (and torques) exerted
by the rest of the monolayer, the patch is submitted to a surface
density  $\bm{w}(x,y)$ of external forces, and to individual external forces acting on the molecules and deriving from a one-body potential energy
$v_\alpha(x,y)$ for the lipid species $\alpha\in\{1,2\}$. The former forces control the shape of the patch and the latter control the mass density distribution of the lipids within the patch. The total potential energy corresponding to these latter forces can be written as 
\begin{equation}
 E_p=\int_{\Omega}\!dn_1v_1+dn_2v_2=\intomeg\psq{\bar\rho\phi\frac{v_1}{\mu_1}+\bar\rho(1-\phi)\frac{v_2}{\mu_2}}\,,
\end{equation}
where $d^2r=dx\,dy$, and $\mu_\alpha$ denotes the mass of one lipid of the species $\alpha$. Introducing $v=v_2/\mu_2$ and $u=v_1/\mu_1-v_2/\mu_2$, this potential energy can be rewritten as: 
\begin{equation}
 E_p=\intomeg\psq{\bar\rho v+\bar\rho\phi u}\,.
\label{ep}
\end{equation}
At equilibrium, the lipid density $\bar\rho$, the composition $\phi$ and the shape $h$ of the monolayer are controlled by the external actions represented by $u(x,y)$, $v(x,y)$ and $\bm{w}(x,y)$.

Let us study a small deformation of the monolayer patch at equilibrium: $\Omega\to\Omega+\delta\Omega$, $h\to h+\delta h$, $\bar\rho\to\bar\rho+\delta\bar\rho$ and $\phi\to\phi+\delta\phi$. Each element of the patch, initially at position $(x,y)$, undergoes a displacement $\delta\bm{a} (x,y)$, with $\delta a_z=\delta h+h_k\delta a_k$ \cite{Fournier07}.
The variation of the free energy of the monolayer patch during the deformation reads
\begin{eqnarray}
\delta F&=&\intomeg\psq{
\deron{\bar f}{\bar\rho}\delta\bar\rho+\deron{\bar f}{\phi}\delta\phi+\deron{\bar f}{\hi}\delta\hi
+\deron{\bar f}{\hij}\delta\hij}\nonumber\\
&+&\intdeltaomeg{\,\bar f},
\end{eqnarray}
We now perform two integrations by parts, and we use the relation 
\begin{equation}
\intdeltaomeg=\intderonomeg ds\,m_i\,\delta a_i\,,
\end{equation}
where $\mathcal{B}$ denotes the boundary of $\Omega$. Assuming that the translation of the monolayer edges is performed at a fixed orientation of its normal, so that $\delta h_j=-h_{jk}\delta a_k$ along the boundary~\cite{Fournier07}, we obtain
\begin{eqnarray}
\delta F&=&\intomeg\psq{
\deron{\bar f}{\bar\rho}\delta\bar\rho+\deron{\bar f}{\phi}\delta\phi+\fder{F}{h}\delta h}\nonumber\\
&+&\intderonomeg ds\,m_i \, \left\lbrace \bar f\delta a_{i}+\psq{\deron{\bar f}{\hi}-\partial_j\deron{\bar f}{\hij}}\delta a_z\right.\\
&+&\left.\psq{\p{\partial_j\deron{\bar f}{\hij}-\deron{\bar f}{\hi}}h_k-\deron{\bar f}{\hij}h_{jk}}\delta a_k\right\rbrace\,\!\!,\nonumber
\label{deltaF}
\end{eqnarray}
where
\begin{equation}
\fder{F}{h}=\partial_k\partial_j\deron{\bar f}{h_{jk}}-\partial_j\deron{\bar f}{h_j}\,.
\end{equation}
The potential energy variation during the deformation is
\begin{eqnarray}
\delta E_p&=&\intomeg\psq{\p{v+u\phi}\delta\bar\rho+u\bar\rho\,\delta\phi}\nonumber\\
&+&\intderonomeg ds\,m_i \,\delta a_i \psq{v\bar\rho+u\bar\rho\phi}\,.
\end{eqnarray}
The total variation $\delta F+\delta E_p$ of the energy of the system must be balanced by the work $\delta W$ done by the surface force density $\bm{w}$ and by the boundary forces exerted by the rest of the membrane on our patch. Since the translation of the monolayer edges is performed at a fixed orientation of its normal, the torques produce no work. We may write
\begin{eqnarray}
\delta W&=&\intomeg\psq{w_k\delta a_k+w_z\delta a _z}\nonumber\\
&+&\intderonomeg ds\,m_i \psq{\Sigma_{ki}\delta a_k+\Sigma_{zi}\delta a_z} \,.
\end{eqnarray}
As the monolayer patch is considered as a closed system, the total mass of each lipid species in the patch is constant during our deformation:
\begin{equation}
\intomeg\,\bar\rho=M\,\,\,\mathrm{and}\,\,\intomeg\,\bar\rho\phi=M_1,
\label{totalmass}
\end{equation}
where $M$ and $M_1$ are constants. Let us introduce two constant Lagrange multipliers $\lambda$ and $\mu$ to implement these two global constraints. The relation $\delta F+\delta E_p-\delta W+\lambda\delta M+\mu\delta M_1=0$ must hold for any infinitesimal deformation of the monolayer patch. The identification of bulk terms in this relation yields
\begin{eqnarray}
w_z&=&\fder{F}{h}\,\,\mathrm{and}\,\,w_k=-h_k w_z\,, \label{rel1}\\
\deron{\bar f}{\bar\rho}&=&-(v+\lambda)-(u+\mu)\phi\,, \label{rel2}\\
\deron{\bar f}{\phi}&=&-(u+\mu)\bar\rho\,\label{rel3}.
\end{eqnarray}
By identifying the boundary terms and using (\ref{rel2}), we obtain the components of the membrane stress tensor: 
\begin{eqnarray}
\Sigma_{ij}&=&\p{\bar f-\bar\rho\deron{\bar f}{\bar\rho}}\delta_{ij}
-\p{\deron{\bar f}{h_j}-\partial_k\deron{\bar f}{h_{kj}}}\hi
\nonumber\\
&-&\deron{\bar f}{h_{kj}}h_{ki},
\label{Sigmaij}
\\
\Sigma_{zj}&=&\deron{\bar f}{h_j}-\partial_k\deron{\bar f}{h_{kj}},
\label{Sigmazj}
\end{eqnarray}
which generalizes the result of Ref.~\cite{Fournier07} to the case where there are inhomogeneities in $\rho$ and $\phi$. Note that the fraction of each lipid species does not appear explicitly in this result. Therefore, Eqs. (\ref{Sigmaij})--(\ref{Sigmazj}) hold both for one-component monolayers and for two-component monolayers. Note also that $\mathbf{\Sigma}$ does not depend directly on the external actions, which confirms its intrinsic nature. 

Comparing our result with Ref.~\cite{Fournier07} shows that taking into account lipid density variations only changes the isotropic term of the stress tensor, which now reads $\bar f-\bar\rho\,\partial \bar f/\partial\bar\rho$. This term is reminiscent of minus the pressure of a two-dimensional homogeneous fluid binary mixture with area $A$ described by a free energy $F(T,A,N_1,N_2)=A  f(T,\rho,\phi)$: 
\begin{equation}
-P=\left.\deron{F}{A}\right|_{T,N_1,N_2}=f-\rho\left.\deron{f}{\rho}\right|_{T,\phi}\,.
\label{Pression}
\end{equation}
While the last expression can be used locally in a non-homogeneous fluid mixture, the case of the membrane is more complex since its free energy depends on the curvature. The ``surface pressure'' in a membrane is sometimes defined as $\rho\,\partial g/\partial\rho -g$ where $g$ is the part of the membrane free-energy density $f$ that depends only on $\rho$ \cite{Cai95} or, equivalently, what remains of $f$ for a planar membrane \cite{Miao02}. Interestingly, we find that the isotropic part of the membrane stress tensor does not identify to minus this surface pressure, since it is the complete, curvature-dependent projected free-energy density $\bar f$ that appears in $\bar f-\bar\rho\,\partial \bar f/\partial\bar\rho$.

The divergence of the stress tensor gives the force per unit area, $\bm{p}$, exerted by the rest of the monolayer on the patch. By direct differentiation, we obtain:
\begin{eqnarray}
p_z&=&\partial_j\Sigma_{zj}=-\fder{F}{h}
\label{pz}
,\\
p_i&=&\partial_j\Sigma_{ij}
=\fder{F}{h}\hi-\bar\rho\,\partial_i\deron{\bar f}{\bar\rho}+\deron{\bar f}{\phi}\partial_i \phi,
\label{pi}
\end{eqnarray}
where we have used
$\partial_if=(\partial \bar f/\partial\bar\rho)\partial_i\bar\rho+(\partial \bar f/\partial\phi)\partial_i\phi
+(\partial \bar f/\partial h_j)\hij+(\partial \bar f/\partial h_{jk}){h_{ijk}}$.
At equilibrium, we can use (\ref{rel1})--(\ref{rel3}) to express $\bm{p}$, which yields
\begin{eqnarray}
p_z&=&-w_z
,\\
p_i&=&-w_i+\bar\rho\partial_i v+\bar\rho\phi\partial_i u\,.
\end{eqnarray}
These relations constitute the balance of surface force densities for the monolayer at equilibrium. In particular, $\bm{p}$ vanishes at equilibrium when the membrane is submitted to no external actions (i.e. $\bm{w}=\mathbf{0}$ and $u=v=0$). 

\section{Monolayer model}
\label{monolayer}

\subsection{Free-energy density in terms of local variables}

\subsubsection{One-component monolayer}

Let us derive the elastic free energy of a monolayer in a bilayer from basic principles, first for a one-component monolayer. We shall recover and extend the model of Ref.~\cite{Seifert93}, which is actually a local version of the ADE model.

We assume that the free-energy density $f$ per unit area of the monolayer depends only on the mass density $\rho$ and on the local principal curvatures $c_1$ and $c_2$ of this monolayer. As in the previous section, the gradients of the curvature and of the density are neglected in our description. Note that, unlike $\bar f$ and $\bar\rho$, $f$ and $\rho$ are the free energy and the mass per actual unit area of the monolayer, and not per projected unit area. We use for both monolayers the density and the principal curvatures defined on the same surface $\mathcal{S}$ of the bilayer, so that the curvatures are common to the two monolayers. 

We will consider the physically relevant regime of curvature radii much larger than the membrane thickness. We will also restrict ourselves to small variations of the density around a reference density $\rho_0$. Note that it can be convenient to take $\rho_0$ different from the equilibrium density $\rho_\mathrm{eq}$ of a plane monolayer with fixed total mass, for instance to study a monolayer under tension. Let us define
\begin{eqnarray}
r&=&\frac{\rho-\rho_0}{\rho_0}=\mathcal{O}\p{\epsilon},\\
H&=&\p{c_1+c_2}\,e=\mathcal{O}\p{\epsilon},\label{H}\\
K&=&c_1c_2\,e^2=\OO\label{K},
\end{eqnarray}
where $e$ is a small length in the nanometer range that allows to define the scaled total curvature $H$ and the scaled Gaussian curvature $K$. 
Since we typically expect $10^{-4}\le|r|\le10^{-2}$ and $10^{-4}\le|c_i\,e|\le10^{-2}$, it is sensible to assume that $r$ and $H$ are $\Oe$ while $K$ is $\OO$. The free-energy density $f$ is a function of these three non-dimensional small variables.

To study small deformations, we write a second-order expansion of $f$:
\begin{eqnarray}
f(r,H,K) &=& \sigma_0+A_1\,H+A_2\p{r-H}^2
+A_3\,H^2 
\nonumber\\
&+& A_4\,K + \OOO.
\label{eqg}
\end{eqnarray}
Three comments are due here. i)We have not included any term linear in $r$ in this expansion. Indeed, the total mass of the monolayer, i.e. the integral of $\rho=\rho_0(1+r)$, is assumed to be constant, so including a term linear in $r$ is equivalent to redefining the constant term $\sigma_0$. ii) The freedom associated with the choice of $e$ allows to set the coefficient of $-rH$ equal to twice that of $r^2$. iii) All the coefficients in (\ref{eqg}) depend on the reference density $\rho_0$. We shall come back in the following on the constitutive relation $\sigma_0(\rho_0)$.

Defining the constants $\kappa$, $\bar\kappa$, $c_0$ and $k$ through
\begin{equation}
A_1=-\frac{\kappa c_0}{2e},\,\,\,\,
A_2=\frac{1}{2}k,\,\,\,\,
A_3=\frac{\kappa}{4e^2},\,\,\,\,
A_4=\frac{\bar\kappa}{2e^2},
\end{equation}
and setting $c=c_1+c_2$, we obtain the following expression, which generalizes those of Refs.~\cite{Seifert93,Miao02}:
\begin{equation}
f = \sigma_0+\frac{k}{2}\p{r-ec}^2
-\frac{\kappa c_0}{2}c+\frac{\kappa}{4}c^2+\frac{\bar\kappa}{2}c_1c_2+
\OOO.
\label{gbo}
\end{equation}
Note that all the above terms have the same order of magnitude. Indeed, since typically $k\approx10^{-1}\,\mathrm{J}/\mathrm{m}^2$, $\kappa\approx\bar\kappa\approx10^{-19}\,\mathrm{J}$, $e\approx1\,\mathrm{nm}$ and $c_0^{-1}\approx50\,\mathrm{nm}$ (see, e.g., Refs.~\cite{Mouritsen_book,Evans90,Safran_book}), we have $A_2\approx A_3\approx A_4\approx100\,A_1$. The advantage of the procedure we have employed is that we control precisely the order of the expansion.

In (\ref{gbo}), $e$ can be interpreted as the distance between the surface $\mathcal{S}$ where $c_1$, $c_2$ and $r$ are defined and the neutral surface of the monolayer~\cite{Safran_book,Petrov84}. As a matter of fact, the density on a surface parallel to $\mathcal{S}$ can be expressed as a function of the distance $\ell$ between $\mathcal{S}$ and this surface as $r(\ell)=r\pm\ell c +\mathcal{O}(\ell c)^2$, where the sign depends on the orientation. We choose the minus sign here, keeping in mind that the second monolayer then has the plus sign (see figure \ref{fig}). Let us now consider the surface such that $\ell=e$. If $f$ is written as a function of $r(e)\equiv r_n$ and of the curvatures, it features no coupling between these variables. This corresponds to the definition of the neutral surface~\cite{Safran_book}, which means that $e$ is the distance between $\mathcal{S}$ and the neutral surface of the monolayer. On this surface, the density which minimizes $f$ for any given membrane shape is $\rho_0$ (at first order in $\epsilon$).

\begin{figure}[htb]
  \begin{center}
    \includegraphics[width=0.75\columnwidth]{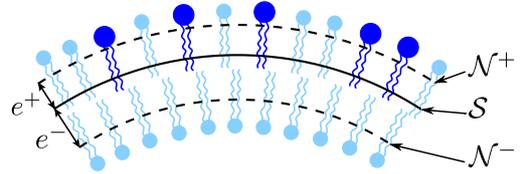}
    \caption{Schematic drawing of a lipid bilayer. The principal curvatures $c_1$, $c_2$ and the scaled densities $r^{\pm}$ of both monolayers are defined on $\mathcal{S}$. The distances between $\mathcal{S}$ and the neutral surfaces $\mathcal{N}^\pm$ of monolayers $\pm$ are denoted $e^\pm$. If the orientation convention is chosen in such a way that $c<0$ on the drawing, the densities on $\mathcal{N}^\pm$ are $r_n^\pm=r^\pm \pm e^\pm c+\OO$. In this example, monolayer $+$ is constituted of two different lipid species.}
  \label{fig}
  \end{center}      
\end{figure}

Let us examine the case of a plane monolayer with fixed total mass. Its equilibrium density $\rho_\mathrm{eq}=\rho_0\p{1+r_\mathrm{eq}}$ can be obtained by minimizing its free energy per unit mass
\begin{equation}
\frac{f}{\rho}=\frac{\sigma_0}{\rho_0}(1-r)+\frac{k+2\sigma_0}{2\rho_0}r^2+\OOO
\end{equation} 
with respect to $r$. We obtain 
\begin{equation}
r_\mathrm{eq}=\frac{\sigma_0}{k+2\sigma_0}\,,\label{r_eq}
\end{equation}
and thus
\begin{equation}
\label{phi}
\sigma_0\p{\rho_0}= k\,\frac{\rho_\mathrm{eq}-\rho_0}{\rho_\mathrm{eq}}+\mathcal{O}\p{\frac{\rho_\mathrm{eq}-\rho_0}{\rho_\mathrm{eq}}}^2.
\end{equation}
In particular, $\sigma_0$ vanishes when $\rho_0=\rho_\mathrm{eq}$.
Note that we have assumed that our reference density $\rho_0$ was sufficiently close to $\rho_\mathrm{eq}$ for the second-order expansion of $f$ to be valid at $\rho_\mathrm{eq}$. 

\subsubsection{Two-component monolayer}

Let us now consider the case of a two-component monolayer. The mass fraction $\phi$ of one lipid species must be taken into accout in our monolayer model. In order to study small variations of $\phi$ around a reference value $\phi_0$, we introduce a fourth small variable 
\begin{equation}
\psi=\frac{\phi-\phi_0}{\phi_0}=\mathcal{O}(\epsilon)\,.
\end{equation}
The expansion of $f$ can now be written as:
\begin{eqnarray}
f &=& \sigma_0+\sigma_1\psi+\frac{\sigma_2}{2}\psi^2+\frac{k}{2}\p{r-ec}^2
-\frac{\kappa}{2} \p{c_0+\tilde{c}_0\psi}c\nonumber\\
&+&\frac{\kappa}{4}c^2+\frac{\bar\kappa}{2}c_1c_2+\OOO\,,
\label{gbo_psi}
\end{eqnarray}
where all the coefficients depend on $\phi_0$ as well as on $\rho_0$.
It is not necessary to include a term in $r\psi$ in this expansion. As a matter of fact, the conservation of the total mass of each lipid species in the monolayer entails that the integral of $\rho\phi$ is a constant as well as the one of $\rho$, so a linear term in $r\psi$ would be redundant with the one in $\psi$.

The equilibrium density $\rho_\mathrm{eq}$ for a flat monolayer with a fixed total mass and a uniform lipid composition $\phi=\phi_0$ has the same expression (\ref{r_eq}) as in the case of a single-component monolayer, but its value depends on $\phi_0$ since $\sigma_0$ and $k$ do.

Expanding $f$ in terms of small variables relies on the assumption that $f$ is analytical. However, in the case where $\phi$ is very small, the free energy contains a non-analytic part, which reads per unit surface $\tilde{\sigma}(1+r)\phi\ln\phi$, with $\tilde{\sigma}=k_\mathrm{B}T\rho_0 /\mu_1$, where $\mu_1$ is the mass of one lipid of the species with mass fraction $\phi$ (see, e.g., the entropy of mixing in Ref.~\cite{Doi_book}). In order to include this term, let us write $f$ as:
\begin{eqnarray}
f &=& \sigma_0+\sigma_1\psi+\frac{\sigma_2}{2}\psi^2+\llbracket\tilde{\sigma}\p{1+r}\phi\ln\phi\rrbracket+\frac{k}{2}\p{r-ec}^2
\nonumber\\&-&\frac{\kappa}{2} \p{c_0+\tilde{c}_0\psi}c+\frac{\kappa}{4}c^2+\frac{\bar\kappa}{2}c_1c_2+\OOO\,,
\label{gbo_phi}
\end{eqnarray}
where the term between double square brackets must be taken into account only if $\phi$ is very small, in which case $\psi$ stands for $\phi$ and not for $(\phi-\phi_0)/\phi_0$. In the following, the double square brackets will always be used with this meaning.

\subsection{Consistency with the ADE model}
\label{Sec:ADE}
 
The area-difference elasticity (ADE) model~\cite{Svetina85,Miao94,Seifert_book} can be deduced from this model by considering a membrane made of two one-component monolayers (denoted $+$ and~$-$ as on figure \ref{fig}) with fixed total masses $M^\pm$, and by eliminating the densities by minimization~\cite{Miao94}. Choosing $\rho_0^\pm=\rho_\mathrm{eq}^\pm$, which implies $\sigma_0^\pm=0$, we have
\begin{equation}
f^\pm = \frac{k^\pm}{2}\p{r^\pm\pm e^\pm c}^2
\pm\frac{\kappa^\pm c_0^\pm}{2}c+\frac{\kappa^\pm}{4}c^2+\frac{\bar\kappa^\pm}{2}c_1c_2\,.
\end{equation}
Since $c_1$ and $c_2$ are defined on the same surface $\mathcal{S}$ of the bilayer, and since a single orientation convention is adopted for both monolayers, the curvatures are common to the two monolayers. Here $e^+$ (resp. $e^-$) is the (positive) distance between $\mathcal{S}$ and the neutral surface of bilayer $+$ (resp. $-$). The $\pm$ sign in front of $c_0^\pm$ ensures that two identical monolayers forming a bilayer would share the same value of $c_0$. Note that the free energy density $f$ defined in the previous section corresponds to $f^-$.

Let us minimize with respect to $r^\pm$ the free energy of each monolayer with fixed total mass $M^\pm$ and total area $A$ (defined on $\mathcal{S}$, like $r^\pm$ and $c_1$ and $c_2$). To take into account the constraints associated with the total masses, we introduce two Lagrange multipliers $\lambda^\pm$. We thus minimize $f^\pm-\lambda^\pm\rho_0^\pm(1+r^\pm)$ with respect to $r^\pm$, which gives 
\begin{equation}
\label{edade}
r^\pm=\frac{\lambda^\pm \rho_0^\pm}{k^\pm}\mp e^\pm c\,.
\end{equation}
Using the constraint $\int_A\!dA\,\rho_0^\pm(1+r^\pm)=M^\pm$ yields
\begin{equation}
\label{constraint}
\frac{\lambda^\pm \rho_0^\pm}{k^\pm}=\frac{M^\pm}{\rho_0^\pm A}
-1\pm\frac{e^\pm}{A}\int_A\!dA\,c\,.
\end{equation}

Let us define the relaxed area $A_0^\pm$ of a monolayer as the area (defined on $\mathcal{S}$) that it would spontaneously adopt in the absence of any constraint:
\begin{equation}
A_0^\pm=M^\pm/\rho_0^\pm\,,
\end{equation}
and its actual area $A^\pm$ measured on the neutral surface of the monolayer, which verifies:
\begin{equation}
\frac{A^\pm}{A}=1\mp \frac{e^\pm}{A} \int_A\!dA\,c +\OO\,.
\end{equation}
We may then rewrite (\ref{constraint}) as $\lambda^\pm \rho_0^\pm/k^\pm=(A_0^\pm-A^\pm)/A$, so that (\ref{edade}) becomes
\begin{equation}
\label{edade2}
r^\pm\pm e^\pm c=\frac{A_0^\pm-A^\pm}{A}\,.
\end{equation}
Thus, after this partial minimization with respect to $r^\pm$, the monolayer free-energy density can be written as
\begin{equation}
f^\pm = \frac{k^\pm}{2}\p{\frac{A^\pm-A_0^\pm}{A}}^2
\pm\frac{\kappa^\pm c_0^\pm}{2}c+\frac{\kappa^\pm}{4}c^2+\frac{\bar\kappa^\pm}{2}c_1c_2\,.\!\!\!
\end{equation}

The total free energy $F^b$ of the bilayer is obtained by integrating $f^++f^-$ over $\mathcal{S}$. Defining the bilayer elastic constants and spontaneous curvature as
\begin{equation}
\kappa^b=\frac{\kappa^+ +\kappa^-}{2},\,\,\, \bar{\kappa}^b=\frac{\bar{\kappa}^+ +\bar{\kappa}^-}{2},\,\,\, c_0^b=\frac{\kappa^- c_0^- -\kappa^+ c_0^+}{\kappa^+ +\kappa^-},\!\!\!
\end{equation}
we obtain
\begin{equation}
F^b=F^b_A + \frac{\kappa^b}{2}\!\int_A\!dA\,c^2 -\kappa^b c_0^b \! \int_A\!dA\,c + \bar{\kappa}^b\!\!\int_A\!dA\,c_1 c_2\,,
\label{F^b}
\end{equation}
where $F^b_A=[k^+ (A^+ -A_0^+)^2+k^- (A^- -A_0^-)^2]/(2A)$.
If the position of $\mathcal{S}$ in the bilayer, which was arbitrary until now, is chosen in such a way that 
\begin{equation}
e^- k^-=e^+ k^+\,,
\label{nsurf}
\end{equation}
$F^b_A$ can be written as
\begin{equation}
F^b_A=\frac{k^b}{A} \p{A-A_0}^2+\frac{K^b}{4A}\p{\Delta A- \Delta A_0}^2\,,
\label{F^bA}
\end{equation}
with $\Delta A=A^- - A^+$, $\Delta A_0=A_0^- - A_0^+$, and
\begin{equation}
k^b=\frac{k^+ +k^-}{2},\,\,\, K^b=\frac{2 k^+ k^-}{k^+ +k^-},\,\,\, A_0=\frac{k^+ A_0^+ +k^-A_0^-}{k^+ +k^-}.\!\!\!
\end{equation}
Condition (\ref{nsurf}) expresses that $\mathcal{S}$ is the neutral surface of the bilayer \cite{Evans74}. Thus, $e^+$ (resp. $e^-$) is the distance between the neutral surface of the monolayer $+$ (resp. $-$) and the neutral surface of the bilayer. 

The total free energy $F^b$ of the bilayer (\ref{F^b})--(\ref{F^bA}), obtained from our monolayer free-energy density, corresponds to the ADE model~\cite{Svetina85,Miao94}.
Note that one usually replaces the term $\propto\!(A-A_0)^2$ by a hard constraint $A=A_0$ \cite{Seifert_book}, while it is important to keep the area-difference elasticity term $\propto\!(\Delta A-\Delta A_0)^2$.

\section{Projected stress tensor}

\subsection{Stress tensor of a monolayer}
In order to obtain the projected stress tensor of our monolayer, we need to determine the projected energy density $\bar f(\bar r,\psi,\hi,\hij)$ associated with $f(r,\psi,c_1,c_2)$. We now assume that the monolayer exhibits only weak deviations from the plane $(x,y)$. Then, if $h_i=\Oe$ and $\hij=\Oe$, Eqs.~(\ref{H})--(\ref{K}) are satisfied because $c_1+c_2=c=\nabla^2h+\OOO$ and $c_1c_2=\det(\hij)+\mathcal{O}(\epsilon^4)$. 

The projected free-energy density $\bar f$ and the projected mass density $\bar\rho$ are given by
\begin{eqnarray}
\bar f&=&f\sqrt{1+\ghc}=f+\frac{\sigma_0}{2}\ghc+\OOO,
\\
\bar\rho&=&\rho\sqrt{1+\ghc}=\rho+\frac{\rho_0}{2}\ghc
+\OOO,\quad
\end{eqnarray}
Hence, defining the scaled projected density as
\begin{equation}
\bar r=\frac{\bar\rho-\rho_0}{\rho_0}=r+\frac{1}{2}\ghc+\OOO,
\label{scaled}
\end{equation}
we obtain from Eq.~(\ref{gbo_phi}) 
\begin{eqnarray}
\bar f&=&\sigma_0+\sigma_1 \psi+\frac{\sigma_2}{2}\psi^2+\frac{\sigma_0}{2}\ghc+\frac{k}{2}\p{\bar r-e\nabla^2h}^2\nonumber\\
&+&\frac{\kappa}{4}\p{\nabla^2h}^2
-\frac{\kappa}{2}\p{c_0+\tilde{c}_0\psi}\nabla^2h
+\frac{\bar\kappa}{2}\det(\hij)\nonumber\\&+&\llbracket\tilde{\sigma}\p{1+\bar r}\phi \ln \phi\rrbracket+\OOO\,.
\label{free}
\end{eqnarray}

We can now calculate the components of the projected stress tensor from (\ref{Sigmaij}) and~(\ref{Sigmazj}), noting that $\bar\rho\,\partial \bar f/\partial\bar\rho=\p{1+\bar r}\partial \bar f/\partial\bar r$. We will restrict ourselves to the first order in $\epsilon$ here, but the tangential components of the stress tensor are calculated at second order in Appendix~\ref{Sig_ordre2}.
We obtain
\begin{eqnarray}
\Sigma_{ij}&=&\psq{\sigma_0+\sigma_1 \psi-k\p{\bar r-e\nabla^2h}-\frac{\kappa c_0}{2}\nabla^2h}\delta_{ij}
\nonumber\\
&+&\frac{\kappa c_0}{2}\hij+\OO,
\label{Sigmaij1}\\
\Sigma_{zj}&=&\sigma_0h_j + ke\,\partial_j\p{\bar r-e\nabla^2h}
-\frac{\kappa}{2}\partial_j\nabla^2h\nonumber\\&+&\frac{\kappa \tilde{c}_0}{2}\partial_j \psi +\OO.
\label{Sigmazj1}
\end{eqnarray}
Note that the non-analytic term coming from the entropy of a mixture, that we have put between double square brackets in (\ref{free}), does not contribute to the stress tensor. This is because this term is proportional to $\bar\rho$, so it disappears when one computes $\bar f-\bar\rho\,\partial \bar f/\partial\bar\rho$. The expression of the stress tensor is thus the same whether $\phi$ is small or whether it is close to a finite value $\phi_0$.

For $h=0$, $\bar r=0$ and $\psi=0$, (\ref{Sigmaij1}) gives $\Sigma_{ij}=\sigma_0\,\delta_{ij}$. Hence $\sigma_0$ can be interpreted as the tension of a flat membrane with uniform density $\rho_0$ and uniform lipid composition $\phi_0$. It is consistent with the fact that $\sigma_0$ vanishes for $\rho_0=\rho_\mathrm{eq}$. Besides, (\ref{phi}) may now be interpreted as a Hookean law for the tension of a flat membrane with no inhomogeneities \cite{Shkulipa06}.

\subsection{Comparison with the stress tensor associated with the Helfrich model}

Let us compare our results with those coming from the Helfrich model, in which the tension $\sigma$ is a phenomenological parameter, namely the Lagrange multiplier implementing the area constraint. For a monolayer with elastic constants $\frac{1}{2}\kappa$ and $\frac{1}{2}\bar\kappa$, tension $\sigma$ and spontaneous curvature $c_0$, the Helfrich free energy density is $f=\sigma+\kappa c^2/4-\kappa c_0 c/2+\bar\kappa c_1c_2/2$, so that its projected version reads at second order in $\epsilon$:
\begin{equation}
\bar f=\sigma+\frac{\sigma}{2}\p{\bm{\nabla}h}^2+\frac{\kappa}{4}\p{\nabla^2 h}^2-\frac{\kappa c_0}{2}\nabla^2h+\frac{\bar\kappa}{2} \det (h_{ij}).\!
\end{equation}
The corresponding stress tensor takes the form~\cite{Fournier07}:
\begin{eqnarray}
\Sigma_{ij}^\mathrm{H}&=&\p{\sigma-\frac{\kappa c_0}{2}\nabla^2h}\delta_{ij}
+\frac{\kappa c_0}{2}\hij+\OO,
\label{SijH1}\\
\Sigma_{zj}^\mathrm{H}&=&\sigma h_j-\frac{\kappa}{2}\partial_j\nabla^2h+\OO.
\label{SzjH1}
\end{eqnarray}

Comparing Eqs.~(\ref{SijH1})--(\ref{SzjH1}) with Eqs.~(\ref{Sigmaij1})--(\ref{Sigmazj1}), we find that we may write
\begin{eqnarray}
\Sigma_{ij}&=&\Sigma_{ij}^\mathrm{H}+\OO,
\label{Sij1}
\\
\Sigma_{zj}&=&\Sigma_{zj}^\mathrm{H}-e\,\partial_j\sigma+\p{\frac{\kappa \tilde{c}_0}{2}+e\sigma_1}\partial_j \psi+\OO,\,\,\,\,
\end{eqnarray}
if we define
\begin{eqnarray}
\sigma&=&\sigma_0+\sigma_1\psi-k\p{\bar r-e\nabla^2h}+\OO.
\end{eqnarray}

Thus, if the scaled lipid composition $\psi$ and the scaled density on the monolayer neutral surface $\bar r_n=\bar r-e\nabla^2 h+\OO$ are both homogeneous, $\sigma$ is a constant. The stress tensor then has the same form in our model as in the Helfrich model. But contrary to the Helfrich tension, our $\sigma$, which may be viewed as a \textit{dynamical surface tension}, can feature inhomogenities. In the inhomogeneous case, new terms appear in $\Sigma_{zj}$. Thus, our stress tensor extends the one associated with the Helfrich model~\cite{Fournier07} to the case where the lipid density and composition are not homogeneous. 

\subsection{Stress tensor in the ADE model}

For a one-component monolayer, the components of the stress tensor at first order in $\epsilon$ (\ref{Sigmaij1})--(\ref{Sigmazj1}) are explicitly given by
\begin{eqnarray}
\Sigma_{xx}&=&\sigma_0-k\psq{\bar r-e\p{h_{xx}+h_{yy}}}
-\frac{\kappa c_0}{2}h_{yy}\,,\\
\Sigma_{xy}&=&\frac{\kappa c_0}{2}h_{xy}\,,
\\
\Sigma_{zx}&=&\sigma_0h_x+ke\,\bar r_x
-\frac{\tilde\kappa}{2}\p{h_{xxx}+h_{xyy}},
\end{eqnarray}
where $\bar r_i\equiv\partial_i\bar r$ and $\tilde\kappa=\kappa+2ke^2$. The other three components follow from exchanging $x$ and $y$.

Even when the membrane exhibits large-scale deformations, it is possible to express the stress tensor at a given point M in the local tangent frame $(X,Y)$ diagonalizing the curvature tensor. Calling $X$ (resp. $Y$) the principal direction associated with the principal curvature $c_1$ (resp. $c_2$), we have $h_X=h_Y=h_{XY}=0$, $h_{XX}= c_1$ and $h_{YY}=c_2$ at point M. Hence $\bar r= r$ and $c=\nabla^2h= c_1+c_2$. The components of the projected stress tensor read at first order:
\begin{eqnarray}
\Sigma_{XX}&=&\sigma_0-k\p{r-ec}-\frac{\kappa c_0}{2}\,c_2,\\
\Sigma_{YY}&=&\sigma_0-k\p{r-ec}-\frac{\kappa c_0}{2}\,c_1,\\
\Sigma_{XY}&=&\Sigma_{YX}=0\,,\\
\Sigma_{ZX}&=&-\frac{\kappa}{2}\partial_X c\,.
\end{eqnarray}
The tangential stress tensor is thus diagonal.

Choosing $\rho_0=\rho_\mathrm{eq}$ and using (\ref{edade2}), which comes from partial minimization of the monolayer free energies with respect to $r^\pm$, we obtain at first order for each monolayer
\begin{eqnarray}
\Sigma_{XX}^\pm&=&k^\pm\,\frac{A^\pm-A_0^\pm}{A}\pm\frac{\kappa^\pm c_0^\pm}{2}\,c_2\,,\\
\Sigma_{YY}^\pm&=&k^\pm\,\frac{A^\pm-A_0^\pm}{A}\pm\frac{\kappa^\pm c_0^\pm}{2}\,c_1\,.
\end{eqnarray}
Summing the contributions from the two monolayers, we obtain the stress tensor of a bilayer in the ADE model, still at first order:
\begin{eqnarray}
\Sigma^b_{XX}&=&2 k^b\,\frac{A-A_0}{A}-\kappa^b c_0^b\, c_2\,,\\
\Sigma^b_{YY}&=&2 k^b\,\frac{A-A_0}{A}-\kappa^b c_0^b\, c_1\,,\\
\Sigma^b_{ZX}&=&-\kappa^b\partial_X c\,.
\end{eqnarray}
where $k^b$, $A_0$, $\kappa^b$ and $c_0^b$ are defined in Sec.~\ref{Sec:ADE}. 

In the Helfrich model, the stress tensor of a bilayer with elastic constant $\kappa^b$ and spontaneous curvature $c_0^b$ can be written in the principal tangent frame from (\ref{SijH1}) and (\ref{SzjH1}). It reads at first order in $\epsilon$:
\begin{eqnarray}
\Sigma^\mathrm{H}_{XX}&=&\sigma-\kappa^b c_0^b\, c_2\,,\\
\Sigma^\mathrm{H}_{YY}&=&\sigma-\kappa^b c_0^b\, c_1\,,\\
\Sigma^\mathrm{H}_{XY}&=&\Sigma^\mathrm{H}_{YX}=0\,,\\
\Sigma^\mathrm{H}_{ZX}&=&-\kappa^b\partial_X c\,.
\end{eqnarray}
Thus, the stress tensor in the ADE model has the same form as the one in the Helfrich model, with $\sigma=2 k^b\p{A-A_0}/A$. In light of the previous section, this equivalence is not surprising since in the ADE model, $\phi=0$, and $\bar r_n$ is homogeneous as shown by (\ref{edade2}).

\section{Force density in a monolayer}

\subsection{Calculation from the projected stress tensor}
\label{div}
Now that we have obtained the stress tensor for our monolayer model, we can calculate the corresponding force per unit area $\bm{p}$ by taking the divergence of (\ref{Sigmaij1}) and (\ref{Sigmazj1}): $p_i=\partial_j\Sigma_{ij}$ and $p_z=\partial_j\Sigma_{zj}$. This force per unit area coming from the rest of the monolayer plays an important part in a dynamical description of a membrane. Indeed, its tangential component is a term of the generalized Navier-Stokes equation describing the monolayer, while its normal component is involved in the normal force balance with the external fluid (see, e.g., Seifert and Langer~\cite{Seifert93}).

We obtain at first order in $\epsilon$:
\begin{eqnarray}
p_i&=&-k\,\partial_i\p{\bar r-e\,\nabla^2h}+\sigma_1\partial_i\psi\,,\label{pi_us}\\
p_z&=&\sigma_0\nabla^2 h-\frac{\tilde{\kappa}}{2}\nabla^4 h+k e\,\nabla^2\bar r + \frac{\kappa \tilde{c}_0}{2}\nabla^2\psi\,\label{pz_us}.
\end{eqnarray}
Both of these results give back those of Ref.~\cite{Seifert93} in the particular case of a bilayer constituted of two identical one-component monolayers, if the reference density is $\rho_0=\rho_\mathrm{eq}$. We have thus justified the expression of these force densities from the membrane stress tensor and generalized them.

This force density can also be derived from the general expressions (\ref{pz}) and (\ref{pi}). 
Note that (\ref{pz}) indicates that $p_z=-\delta F/\delta h$, which is indeed the force taken into account in Eq.~(3) of Ref.~\cite{Seifert93}. Besides, applying (\ref{pi}) to the free-energy density (\ref{free}) with $\phi=0$ shows that, in this specific case, $p_i=-\partial_i(\delta F/\delta \bar r)+\OO$. This justifies the ``gradient of the surface pressure" term $-\bm{\nabla}(\delta F/\delta r)$ used in Eq.~(4) of Ref.~\cite{Seifert93}. 

\subsection{Direct covariant calculation}
\label{cov}

\subsubsection{Definitions and notations}
In this section, we will not restrict ourselves to membranes undergoing small deformations around the flat shape. In general, a membrane can be considered, in a coarse-grained description, as a two-dimensional surface embedded in the three-dimensional space. The position of a fluid element in the membrane can be described by a three-dimensional vector $\bm{R}(u^1,u^2)$, where $u^1$ and $u^2$ are two parameters labelling each fluid element. Mathematically, these parameters are internal coordinates in the two-dimensional surface, and physically they correspond to Lagrangian coordinates. 

We are now going to introduce some basic definitions and notations used to describe the shape of a surface in differential geometry \cite{Aris,Cai95,Miao02}.  At each point $\bm{R}$ of the surface, it is possible to define two vectors tangent to the surface through 
\begin{equation}
\bm{t}_{\alpha}=\frac{\partial\bm{R}}{\partial u^\alpha}\equiv\partial_\alpha \bm{R}\,,
\end{equation}
where $\alpha\in\{1,2\}$. These two vectors are supposed to be linearly independent. Thus, 
\begin{equation}
\bm{n}=\frac{\bm{t}_1\times\bm{t}_2}{|\bm{t}_1\times\bm{t}_2|}
\end{equation}
is a unit normal to the surface at point $\bm{R}$. The metric tensor of the surface can be expressed as 
\begin{equation}
a_{\alpha\beta}=\bm{t}_\alpha\cdot\bm{t}_{\beta}\,,
\end{equation}
so that the area element of the surface reads 
\begin{equation}
dA=\sqrt{a}\,d^2u\,, 
\end{equation}
where $a$ is the determinant of $a_{\alpha\beta}$, and $d^2 u=du^1 du^2$. The inverse metric tensor $a^{\alpha\beta}$ is defined by the relation
\begin{equation}
a^{\alpha\beta}a_{\beta\gamma}=\delta^\alpha_\gamma\,, 
\end{equation}
where $\delta^\alpha_\gamma$ is the Kronecker symbol.
In the last relation, as well as in the following, the Einstein summation convention is used. We may now define the contravariant tangent vectors as
\begin{equation}
\bm{t}^\alpha=a^{\alpha\beta}\bm{t}_{\beta}\,. 
\end{equation} 
A complete description of a surface is given by its metric tensor (or first fundamental form) and its curvature tensor (or second fundamental form)
\begin{equation}
b_{\alpha\beta}=\bm{n}\cdot\partial_\alpha\bm{t}_\beta=\bm{n}\cdot \partial_\alpha \partial_\beta \bm{R}\,.
\end{equation}
The principal curvatures $c_1$ and $c_2$ of the surface are the eigenvalues of $b^\alpha_\beta=a^{\alpha\gamma}b_{\gamma\beta}$, which enables to express the total curvature and the Gaussian curvature from the curvature tensor:
\begin{eqnarray}
c&=&c_1+c_2=b^\alpha_\alpha\,,\\
c_1c_2&=&\det b^\alpha_\beta\,.
\end{eqnarray}

\subsubsection{Force density in a monolayer}
The surface density of internal forces $\bm{q}$ in a two-component monolayer with free energy $F=\int\!dA\,f$ can be expressed as the functional derivative
\begin{equation}
\bm{q}(u^1,u^2)=-\frac{1}{\sqrt{a}} \left. \frac{\delta F}{\delta \bm{R}(u^1,u^2)}\right|_{\rho\sqrt{a},\,\,\rho\phi\sqrt{a}}\,,
\label{q}
\end{equation}
where, as in the previous sections, $\rho$ is the total mass density of lipids, and $\phi$ the mass fraction of one lipid species \cite{Cai95,Miao02,Lomholt05}. This expression is a consequence of the principle of virtual work: for a small deformation $\delta\bm{R}$ of the membrane at equilibrium, the membrane free-energy variation reads 
\begin{equation}
\delta F=-\int\! dA \,\,\bm{q}\cdot\delta\bm{R}=-\int\!d^2u \,\,\sqrt{a}\,\,\bm{q}\cdot\delta\bm{R}\,.
\label{pvw}
\end{equation}
For the underlying force balance on each fluid element to be valid, the virtual deformation $\delta\bm{R}$ must be performed at constant total mass $dm=\rho\, dA=\rho\sqrt{a}\,\,d^2u$ and composition in each fluid element. Hence, the functional derivative in (\ref{q}) must be taken at constant $\rho\sqrt{a}$ and $\rho\phi\sqrt{a}$.

Let us calculate the force density (\ref{q}) in a monolayer with free-energy density
\begin{eqnarray}
f &=& \sigma_0+\sigma_1\psi+\frac{\sigma_2}{2}\psi^2+\llbracket\tilde{\sigma}\p{1+r}\phi\ln\phi\rrbracket+\frac{k}{2}\p{r-ec}^2
\nonumber\\&-&\frac{\kappa}{2} \p{c_0+\tilde{c}_0\psi}c+\frac{\kappa}{4}c^2+\frac{\bar\kappa}{2}c_1c_2.
\label{gbo_trunc}
\end{eqnarray}
This free-energy density corresponds to (\ref{gbo_phi}) truncated at second order in $\epsilon$. The two constraints on the deformation $\delta\bm{R}$, $\delta (\rho\sqrt{a})=0$ and $\delta (\rho\phi\sqrt{a})=0$, are equivalent to
\begin{eqnarray}
\sqrt{a} \,\delta r + (1+r)\delta\sqrt{a}&=&0 \label{ctrte_1}\\
\sqrt{a}\, \delta\psi&=&0 \label{ctrte_2}\,,
\end{eqnarray}
where we have used the fact that $1+r>0$.
To enforce these two independent local constraints, we use two local Lagrange multipliers, $\lambda$ and $\mu$. The principle of virtual work (\ref{pvw}) then reads
\begin{equation}
\delta F-\int\!d^2u\left\{\lambda\psq{\sqrt{a}\, \delta r + (1+r)\delta\sqrt{a}}+\mu\,\sqrt{a} \,\delta\psi\right\}=\delta W\,,
\label{virtw}
\end{equation}
where 
\begin{equation}
\delta W=-\int\! dA \,\,\bm{q}\cdot\delta\bm{R}\,.\label{travail}
\end{equation}
We assume that the topology of the membrane is not affected by the virtual deformation. The Gauss-Bonnet theorem then ensures that $\delta\p{\int\!dA\,\,c_1c_2}=0$. The transformation of the left-hand side of (\ref{virtw}) can be performed along the same lines as in Ref.~\cite{Jenkins77}. These calculations, which are presented in Appendix~\ref{Calculq}, yield:
\begin{eqnarray}
\bm{q}\cdot\bm{t}_\alpha&=&-k\p{1+r}\partial_\alpha\p{r-ec}\nonumber\\
&&+\p{\sigma_1+\sigma_2\psi-\frac{\kappa\tilde c_0}{2}c}\partial_\alpha\psi\,,\label{q_tg}\\
\bm{q}\cdot\bm{n}&=&\p{\sigma_0+\sigma_1\psi+\frac{\sigma_2}{2}\psi^2-\frac{k}{2}r^2-k r}c-\frac{\tilde\kappa}{4}c^3\nonumber\\
&&+\psq{\tilde\kappa c -\kappa \p{c_0+\tilde c_0 \psi}-2 k e\,\,r}c_1c_2\nonumber\\
&&+ke \p{1+r}c^2+ke\,\Delta r-\frac{\tilde\kappa}{2}\Delta c+\frac{\kappa\tilde c_0}{2}\Delta\psi\,,\quad\quad\label{q_n}
\end{eqnarray}
where $\Delta$ is a shorthand for the Laplace-Beltrami operator $(1/\sqrt{a})\partial_\alpha(a^{\alpha\beta}\sqrt{a} \partial_\beta)$. The force density $\bm{q}$ can be expressed from its tangential component (\ref{q_tg}) and normal component (\ref{q_n}) as 
\begin{equation}
\bm{q}=(\bm{q}\cdot\bm{t}_\alpha)\bm{t}^\alpha+(\bm{q}\cdot\bm{n})\bm{n}\,.\label{qtot}
\end{equation}
We have thus obtained the general expression of the force density in a two-component monolayer with free-energy density (\ref{gbo_trunc}). This expression gives back the one in Ref.~\cite{Miao02} in the particular case of a one-component monolayer with $c_0=0$. 

Note that, in this section, we have used the free-energy density truncated at second order (\ref{gbo_trunc}) as if it were exact. The force density $\bm{q}$ expressed in (\ref{q_tg})--(\ref{qtot}) is the one corresponding to this model, and it contains second and third-order terms. This approach is consistent with the one of Refs.~\cite{Jenkins77,Cai95,Miao02}.  
However, in the present paper, we have constructed the free-energy density as a general expansion around a reference state, controlling the order in $\epsilon$ of this expansion. In our approach, if the free-energy density $f$ is kept at second order, the force density can be known only at first order.

\subsection{Comparison between the two results}

In the present paper, except in the previous section \ref{cov}, we have described membranes in the Monge gauge, i.e., by their height with respect to a reference plane. Such a description is very convenient to study the membrane small deformations around the flat shape. In the Monge gauge, the position of a fluid element in the membrane is given by $\bm{R}(x,y)=(x,\,y,\,h(x,y))$, where $x$ and $y$ are Cartesian coordinates in the reference plane and $z=h(x,y)$ describes the height of the membrane with respect to the reference plane. Obviously, the coordinates $(x,y)$ of a fluid element depend on its position in the membrane (in other words, they are Eulerian coordinates). 

In Sec.~\ref{cov}, we have found the force density $\bm{q}$ in a monolayer, whatever its shape. In our derivation of $\bm{q}$, the parameters $(u^1,u^2)$ describing the surface were Lagrangian coordinates, labelling each fluid element. However, as the force density is a physical quantity, it does not depend on the parametrization of the surface \cite{Cai95}. Thus, the expression we have found for $\bm{q}$ is valid (for each given membrane shape) in the Monge gauge.

We may now compare the force density obtained in Sec.~\ref{cov} with the one obtained from the projected stress tensor in Sec.~\ref{div}. For this, we shall write explicitly in the Monge gauge the general result obtained in Sec.~\ref{cov}. With $u^1=x$ and $u^2=y$, the tangent vectors read in the Monge gauge $\bm{t}_1=(1,\,0,\,\partial_x h)$ and $\bm{t}_2=(0,\,1,\,\partial_y h)$. It is then straightforward to find the expression of $\bm{n}$, $a_{\alpha\beta}$ and $a^{\alpha\beta}$ in the Monge gauge (see, e.g., Ref.~\cite{Cai95}). Using these explicit expressions, and keeping only first order terms in $\epsilon$, (\ref{q_tg}) and (\ref{q_n}) can be written as:
\begin{eqnarray}
q_i&=&-k\,\partial_i\p{\bar r-e\,\nabla^2h}+\sigma_1\partial_i\psi+\OO,\label{qi}\\
q_z&=&\sigma_0\nabla^2 h-\frac{\tilde{\kappa}}{2}\nabla^4 h+k e\,\nabla^2\bar r + \frac{\kappa \tilde{c}_0}{2}\nabla^2\psi+\OO,\label{qz}\quad\quad
\end{eqnarray}
where $i\in\{x,y\}$.
We notice that, at this order, (\ref{qi}) is identical to (\ref{pi_us}) and (\ref{qz}) is identical to (\ref{pz_us}). Note that $\bm{q}$ is a force density per actual unit area of the monolayer while $\bm{p}$ is a force density per projected unit area. However, this difference is irrelevant at first order.

We have just shown that the force density obtained from the divergence of the projected stress tensor is consistent with the one calculated directly by using the principle of virtual work in covariant formalism. The projected stress tensor thus allows to calculate easily both the normal and the tangential components of the force density in a membrane in the Monge gauge, without having to resort to a covariant formulation.

\section{Applications}

The force density in a membrane with lipid density and composition inhomogeneities can be used to understand qualitatively and quantitatively the dynamics of a membrane submitted to a local perturbation. In order to illustrate this, we are going to focus on the local injection of a reagent close to a membrane (see Fig.~\ref{Sit}), which modifies locally the properties of the membrane. 
\begin{figure}[htb]
  \begin{center}
    \includegraphics[width=0.75\columnwidth]{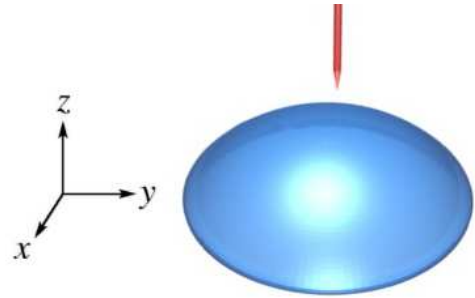}
    \caption{Local injection of a reagent from a micropipette close to a vesicle. The lipids in the external monolayer of the vesicle will be affected. The injection is sufficiently local for us to focus on a small, nearly-plane zone of the membrane.}
  \label{Sit}
  \end{center}      
\end{figure}

\subsection{Forces arising from a modification of composition}
Let us consider initially a one-component flat membrane at equilibrium with uniform lipid density. 
Let us assume that, at time $t=0$, some lipids in the external monolayer of this membrane (monolayer $+$, as on Fig.~\ref{fig}), are suddenly chemically modified due to a microinjection of a reagent close to the membrane (see Fig.~\ref{Sit}). Then there is a fraction $\phi(x,y)$, assumed to be small, of modified lipids in this monolayer. The force density in the external monolayer at time $t=0^+$, just after the injection, when the shape and the density have not changed yet, is given by: 
\begin{eqnarray}
\bm{p}^+&=&\sigma_1\bm{\nabla}\phi+\frac{\kappa \tilde{c}_0}{2}\nabla^2\phi\,\,\bm{e}_z\,,
\end{eqnarray}
where $\bm{e}_z$ is a unit vector in the $z$ direction. This force density corresponds to Eqs.~(\ref{pi_us}) and (\ref{pz_us}) in the case of a flat ``+'' monolayer with uniform density.  
Hence, modifying locally the lipids of a monolayer will generically induce a shape instability of the membrane. 
The internal monolayer is not affected by the chemical modification, so the force density remains zero in it. 

Let us take the position of the micropipette injecting the reagent as the origin of our $(x,y)$ frame. Then, $\phi$ is a decreasing function of the radial coordinate $r$. Let us study the case where $\phi$ is a Gaussian: 
\begin{equation}
\phi(r)=\phi_0\,\exp\p{-\frac{r^2}{2\,R^2}}\,.                                                                                                         \end{equation}
This can represent the field of modified lipids resulting from a diffusion of the reagent in the solution surrounding the vesicle before it hits the membrane.
It is straightforward to calculate the corresponding force density. Its normal and radial components, nondimensionalized by their maximal values, are plotted as a function of $r/R$ in Fig.~\ref{fd1}. 
\begin{figure}[htb]
  \begin{center}
    \includegraphics[width=0.75\columnwidth]{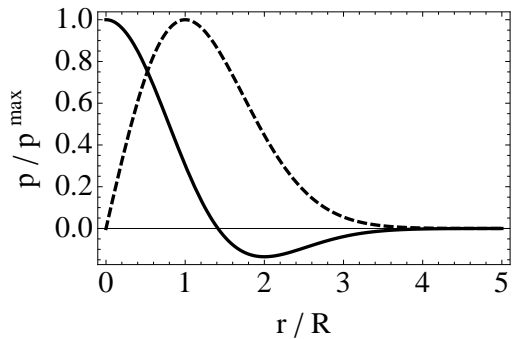}
    \caption{Nondimensionalized force density in the external monolayer of a membrane just after a local chemical modification, with Gaussian $\phi$. Plain line: normal component $p_z/p_z^\mathrm{max}$. Dashed line: radial component $p_r/p_r^\mathrm{max}$.}
  \label{fd1}
  \end{center}      
\end{figure}

The constants $\sigma_1$ and $\tilde c_0$ that appear in the force density arise from the $\phi$-dependence of the free energy per unit area of a monolayer (see Eq.~\ref{gbo_phi}). Physically, they describe the change in the equilibrium density and the spontaneous curvature of the membrane due to a generic modification of the lipids \cite{FuturMig}. Their signs, and thus, those of the force density components, depend on the nature of the modification. In the case where $\sigma_1<0$ and $\tilde c_0<0$, which corresponds to modified lipids favoring a smaller density and a larger curvature (in absolute value), the lipids in front of the pipette are submitted to a normal force going towards the exterior of the vesicle, which will yield a local deformation of the membrane in this direction. 
Meanwhile, a radial force drives the lipids of the external monolayer to flow in the membrane towards larger values of $r$, due to the fact that the modified lipids favor a smaller density. 

This situation describes well the onset of the shape instability studied in Ref.~\cite{FuturMig}, which is induced by microinjecting a basic solution close to a giant unilamellar vesicle. This instability can yield the formation of a membrane tubule \cite{Fournier09}.

\subsection{Forces arising from a local deformation at uniform density}
Besides the composition, another important effect captured by our study is the coupling between the membrane shape and the density. To shed light onto this effect, let us consider a locally deformed membrane with uniform density (on the bilayer midsurface). This can correspond to a membrane which has deformed very rapidly from a flat shape, before the density adjusts to the new deformed shape. Indeed, the symmetric density (i.e., the sum of the densities in the two monolayers) is not coupled to the deformation, while the antisymmetric density is~\cite{Seifert93,Evans94,Fournier09,FuturMig}. Thus, intermonolayer friction is involved when the density adjusts to the deformation. The associated timescale can be quite large, e.g., a few seconds for deformations on length scales of order 20 to 100 $\mu$m, so there is indeed a lapse when a deformed membrane with non-adjusted density exists in the experiments described in Ref.~\cite{FuturMig}.

To isolate the effect of the shape and density, we will focus on a one-component membrane here (note that the external monolayer is a two-component one in Ref.~\cite{FuturMig}). Let us take $\rho_0=\rho_\mathrm{eq}$ as our reference density in both (identical) monolayers. Then, Eq.~(\ref{phi}) ensures that $\sigma_0=0$. The force density in monolayers ``$\pm$'' caused by the deformation is given by: 
\begin{eqnarray}
\bm{p}^\pm&=&\mp ke\,\bm{\nabla}(\nabla^2h)-\frac{\tilde{\kappa}}{2}\nabla^4 h\,\bm{e}_z\,.
\end{eqnarray}
This force density corresponds to Eqs.~(\ref{pi_us}) and (\ref{pz_us}) in the case of one-component monolayers with uniform density. 

Let us consider for instance a Gaussian-shaped deformation towards the exterior, centered on the origin:
\begin{equation}
h(r)=h_0\,\exp\p{-\frac{r^2}{2\,R^2}}\,.
\end{equation}
The normal and radial components of the corresponding force density, nondimensionalized by the absolute value of their maxima, are plotted as a function of $r/R$ in Fig.~\ref{fd2}. 
\begin{figure}[htb]
  \begin{center}
    \includegraphics[width=0.75\columnwidth]{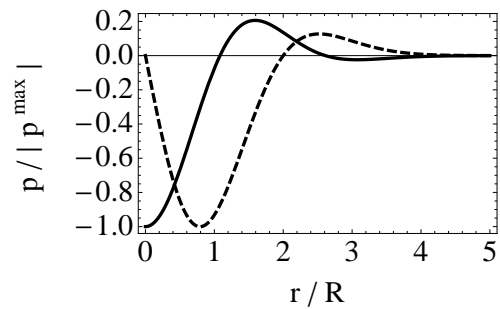}
    \caption{Nondimensionalized force density in monolayer ``$+$'' just after a local deformation, with Gaussian $h$. Plain line: normal component $p_z/|p_z^\mathrm{max}|$. Dashed line: radial component $p_r/|p_r^\mathrm{max}|$. In monolayer ``$-$'', $p_z$ is identical and $p_r$ is opposite.}
  \label{fd2}
  \end{center}      
\end{figure}

Since $ke>0$ and $\tilde\kappa>0$, at small $r$, the lipids are submitted to a normal force going towards the interior of the vesicle: this will lead to a relaxation of the deformation. This is due to the fact that the membrane considered here is symmetric, so its equilibrium shape is flat. 
Meanwhile, a radial force drives the lipids of the external monolayer to flow in the membrane, for the density to adjust to the shape (see Fig.~\ref{dbd}). Indeed, in a curved membrane at equilibrium, the density is uniform on the neutral surface of each monolayer, and not on the membrane midsurface. The radial forces are opposite in the external and in the internal monolayer, because the orientations of these monolayers are opposite while they share the same curvature. Hence, for the density to adjust to the deformed shape, it is necessary that the lipids in one monolayer slide with respect to the ones in the other monolayer, which involves intermonolayer friction \cite{FuturMig}.
\begin{figure}[htb]
  \begin{center}
    \includegraphics[width=0.9\columnwidth]{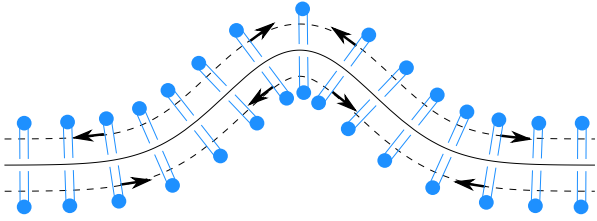}
    \caption{Membrane with Gaussian $h$, that has just deformed from a flat shape. The density has not adjusted to the new deformed shape yet. The lipids are thus at equal distance on the midlayer (plain line). The equilibrium density for this deformed shape would correspond to lipids at equal distance on the neutral surface of each monolayer (dashed lines). The arrows indicate the direction of the tangential force density in the membrane, consistent with Fig.~\ref{fd2}.}
  \label{dbd}
  \end{center}      
\end{figure}

\subsection{Dynamics}
The force densities in the membrane are the basis of a hydrodynamic description of the membrane (for a review, see Ref.~\cite{Powers10}). Our work enables to take into account lipid density and composition instabilities in such dynamical studies. 
In Ref.~\cite{FuturMig}, we have used the force densities derived in the present work to describe the dynamics of a membrane curvature instability induced by a local injection of a basic solution close to a giant unilamellar vesicle.

More precisely, we have written down generalized Stokes equations describing the balance of the forces per unit area acting tangentially in each monolayer. These equations include the tangential density of forces Eq.~(\ref{pi_us}), the viscous force density due to the two-dimensional flow of the lipids, the tangential viscous stress exerted by the flow of the surrounding fluid, and the intermonolayer friction. We have also written down the balance of the forces per unit area acting normally to the membrane, including the normal force density given by Eq.~(\ref{pz_us}), and the normal viscous stress exerted by the flow of the surrounding fluid. Our last fundamental dynamical equation is the conservation of mass. 

Thanks to these equations, we have described theoretically the evolution of the deformation resulting from the local chemical modification of the external monolayer by the basic solution. This description is developed in detail, and compared to experimental results in Ref.~\cite{FuturMig}.

\section{Conclusion}
We have derived a general formula expressing the projected stress tensor in a monolayer as a function of the monolayer free-energy density, taking into account inhomogeneities in the lipid density and composition. This general formula has been applied to a generic monolayer model constructed from basic principles. Our model being consistent with the ADE model, we have found in particular the stress tensor associated with the ADE model.

In the Monge gauge, the projected stress tensor provides a convenient way of deriving the force density in a monolayer, which is the basis of a hydrodynamic description of a membrane. The result is consistent with a direct calculation of the force density from the principle of virtual work in covariant formalism. We have shown an example of application to the calculation of force density in a locally perturbed membrane. These force density constitute the basis of the dynamical study in Ref.~\cite{FuturMig}.

Moreover, the stress tensor contains more information than the force density, since it provides the actual force exerted by a piece of membrane along its edge. Indeed, the stress tensor associated with the Helfrich model \cite{Capovilla02,Fournier07} has already been used to study various situations, such as the boundary conditions on a membrane with a free exposed edge \cite{Capovilla02b}, the adhesion of a fluid membrane \cite{Capovilla02c,Deserno07}, the surface tension of fluctuating membranes \cite{Fournier08,Barbetta10} and the force exerted by a fluctuating membrane tubule \cite{Barbetta09}. The stress tensor is also a useful tool in the study of membrane-mediated interactions \cite{Muller05,Muller05b,Bitbol10}. 
Since the stress tensor studied in the present paper generalizes the one associated with the Helfrich model to monolayers with density and composition inhomogeneities, it may enable to extend such applications. 

\appendix
\section{Tangential stress tensor at second order}
\label{Sig_ordre2}
For the Helfrich model, the tangential components of the stress tensor can be obtained at second order from the free energy at second order\cite{Fournier07}. On the contrary, here, because of the term $\partial \bar f/\partial\bar r$, one cannot obtain $\Sigma_{ij}$ at order $\epsilon^2$ without taking into account in $\bar f$ the terms of order $\epsilon^3$ which depend on $\bar r$.

When such third order terms are included, $f$ becomes
\begin{eqnarray}
f &=& \sigma_0+\sigma_1\psi+\frac{\sigma_2}{2}\psi^2+\llbracket\tilde{\sigma}\p{1+r}\phi\ln\phi\rrbracket\nonumber\\
&+&\frac{k}{2}\p{r-ec}^2-\frac{\kappa}{2} \p{c_0+\tilde{c}_0\psi}c+\frac{\kappa}{4}c^2+\frac{\bar\kappa}{2}c_1c_2 \nonumber\\
&+& \frac{\sigma'_1}{2} r^2\psi+\frac{\sigma'_2}{2}r\psi^2+\frac{k'}{3}r^3 +\frac{\kappa c'_0}{2}r^2 c-\frac{\kappa\tilde{c}'_0}{2}\psi r c\nonumber\\
&+& \frac{\kappa'}{4}r c^2+\frac{\bar\kappa'}{2}rc_1c_2 +\OOOOp,\quad
\end{eqnarray}
where $\OOOOp$ stands for terms of order $\epsilon^4$, \textit{or terms of order $\epsilon^3$ independent of $r$}. Indeed, the latter may be discarded because they will not contribute to the stress tensor at order $\epsilon^2$. The new terms in $f$ may be considered as originating from a density-dependence of the constitutive constants $k$, $\kappa$, $\bar{\kappa}$, $c_0$, $\tilde{c}_0$, $\sigma_1$ and $\sigma_2$.

Using the relations
\begin{eqnarray}
\bar f&=&f\sqrt{1+\ghc}=f+\frac{\sigma_0}{2}\ghc+\OOOOp,\quad\\
c&=&\nabla^2h+\OOOOp,\\
c_1c_2&=&\det(\hij)+\OOOOp,\\
\bar r&=&r+\frac{1}{2}\ghc+\OOOOp,
\end{eqnarray}
we obtain
\begin{eqnarray}
\bar f&=&\sigma_0+\frac{\sigma_0}{2}\ghc+\sigma_1\,\psi+\frac{\sigma_2}{2}\,\psi^2+\llbracket\tilde{\sigma}\p{1+\bar r}\phi\ln\phi\rrbracket\nonumber\\
&+&\frac{k}{2}\p{\bar r-e\nabla^2h}^2 -\frac{\kappa}{2}\p{c_0+\tilde{c}_0\psi}\nabla^2h+\frac{\kappa}{4}\p{\nabla^2h}^2
\nonumber\\
&+&\frac{\bar\kappa}{2}\det(\hij)+\bigg[\frac{\kappa'}{4}\p{\nabla^2h}^2+\frac{\bar\kappa'}{2}\det\p{\hij}\nonumber\\
&-&\frac{k}{2}\p{\bm{\nabla}h}^2+\frac{\sigma'_2}{2}\,\psi^2-\frac{\kappa\tilde{c}'_0}{2}\,\psi\nabla^2 h\bigg]\bar r\nonumber\\
&+&\p{\frac{\sigma'_1}{2}\,\psi-\frac{\kappa c'_0}{2}\,\nabla^2h}\bar r^2 +\frac{k'}{3}\,\bar r ^3+\OOOOp.
\label{free3}
\end{eqnarray}

Eqs.~(\ref{Sigmaij}) and~(\ref{free3}) yield
\begin{eqnarray}
\Sigma_{ij}&=&\bigg\{
\sigma_0+\sigma_1\,\psi+\frac{\sigma_2-\sigma'_2}{2}\,\psi^2+\frac{\sigma_0+k}{2}\ghc\nonumber\\
&&\,\,-\p{k+\sigma'_1\psi}\bar r-\p{k+2k'}\frac{\bar r^2}{2}+\frac{\tilde\kappa-\kappa'}{4}\p{\nabla^2h}^2\nonumber\\
&&\,\,+\psq{ke-\frac{\kappa}{2}\p{c_0+\tilde{c}_0\psi}+\kappa c_0'\,\bar r+\frac{\kappa\tilde{c}'_0}{2}\,\psi}\nabla^2h
\nonumber\\
&&\,\,-\frac{\bar\kappa'}{2}\det\p{\hij}\bigg\}\delta_{ij}-\sigma_0\,h_ih_j
+\frac{\kappa}{2}\p{c_0+\tilde{c}_0\psi}h_{ij}
\nonumber\\
&-&\frac{\tilde\kappa}{2}\p{h_{ij}\nabla^2h-h_i\partial_j\nabla^2h}
+ke\p{h_{ij}\bar r-h_i\partial_j\bar r}\nonumber\\
&-&\frac{\kappa\tilde{c}_0}{2}\,h_i\partial_j\psi+\OOO.
\label{Sijfull}
\end{eqnarray}
where $\tilde\kappa=\kappa+2ke^2$ as before. The non-analytic term present in $\bar f$ at small $\phi$ does not contribute to the stress tensor, for the same reason as before.

In the principal tangent frame, the tangential components of the stress tensor at second order are given by
\begin{eqnarray}
\Sigma_{XX}&=&\sigma_0+\sigma_1\,\psi+\frac{\sigma_2-\sigma'_2}{2}\,\psi^2-\p{k+\sigma'_1\psi} r\nonumber\\
&&-\p{k+2k'}\frac{r^2}{2}+\p{k e +\kappa c'_0\,r+\frac{\kappa\tilde{c}'_0}{2}\,\psi}c\nonumber\\
&&-\frac{\kappa}{2}\p{c_0+\tilde{c}_0\psi}c_2+k e\,\,r c_1\nonumber\\
&&-\frac{\tilde{\kappa}+\kappa'}{4}\,c_1^2+\frac{\tilde{\kappa}-\kappa'}{4}\,c_2^2-\frac{\kappa'+\bar\kappa'}{2}\,c_1c_2,\\
\Sigma_{XY}&=&\Sigma_{YX}=0\,.
\end{eqnarray}
$\Sigma_{YY}$ can be obtained by exchanging $c_1$ and $c_2$ in $\Sigma_{XX}$. This tangential stress tensor thus remains diagonal at second order in the principal tangent frame, like the one associated with the Helfrich model \cite{Fournier07}.

\section{Covariant calculation of the force density}
\label{Calculq}
In this section, we are going to present the main steps of our covariant calculation of the force density in a monolayer, which leads to (\ref{q_tg})--(\ref{q_n}). This calculation follows the same lines as the one in Ref.~\cite{Jenkins77}.

As the Gaussian curvature term contained in the total monolayer free energy
\begin{equation}
F=\int dA\,f=\int\!d^2u\,\sqrt{a}\,f
\end{equation}
does not vary during the virtual deformation $\delta\bm{R}$, we may replace the free-energy density (\ref{gbo_trunc}) by
\begin{equation}
\tilde f=f-\frac{\bar\kappa}{2}c_1 c_2
\end{equation}
in our calculations.
The free-energy variation during the deformation then reads
\begin{equation}
\delta F=\int\!d^2u\,\tilde f\,\delta\sqrt{a}+\int\!d^2u\,\sqrt{a}\,\delta \tilde f\,,
\end{equation}
with
\begin{eqnarray}
\delta \tilde f&=&\psq{\sigma_1+\sigma_2\psi-\frac{\kappa\tilde c_0}{2}c+\left\llbracket\tilde\sigma\p{1+r}\p{1+\ln\phi}\right\rrbracket}\delta\psi\nonumber\\
&+& \big[k(r-ec)+\left\llbracket\tilde\sigma\phi\ln\phi\right\rrbracket\big]\,\delta r+I_1\delta c\,,
\end{eqnarray}
where we have defined
\begin{equation}
I_1=\frac{\kappa}{2}\p{c-c_0-\tilde c_0\psi}-ek\p{r-ec}\,.
\end{equation}

The principle of virtual work (\ref{pvw}) may now be written as
\begin{eqnarray}
\delta W&=&\int\!dA\,\Bigg\{I_1\delta c+\big[k(r-ec)+\left\llbracket\tilde\sigma\phi\ln\phi\right\rrbracket-\lambda\big]\,\delta r\nonumber\\
&+&\psq{\sigma_1+\sigma_2\psi-\frac{\kappa\tilde c_0}{2}c+\left\llbracket\tilde\sigma\p{1+r}\p{1+\ln\phi}\right\rrbracket-\mu}\delta\psi\Bigg\}\nonumber\\
&+&\int\!d^2 u\psq{\tilde f -\lambda\p{1+r}}\delta\sqrt{a}\,,
\label{trav_app}
\end{eqnarray}
where $\delta W$ is given by (\ref{travail}). The variations $\delta c$ and $\delta\sqrt{a}$ only come from the variation $\delta\bm{R}$ of the \emph{shape} of the monolayer. The coupling between  $\delta\bm{R}$ and $\delta r$ and $\delta \psi$, which comes from the constraints (\ref{ctrte_1})--(\ref{ctrte_2}), has been accounted for by introducing the Lagrange multipliers $\lambda$ and $\mu$. Hence, $\delta \bm{R}$, $\delta r$ and $\delta \psi$ should now be considered as independent variations, and the terms in $\delta r$ and $\delta \psi$ must vanish for (\ref{trav_app}) to be valid for any virtual deformation, yielding
\begin{eqnarray}
\lambda&=&k(r-ec)+\left\llbracket\tilde\sigma\phi\ln\phi\right\rrbracket\,,\\
\mu&=&\sigma_1+\sigma_2\psi-\frac{\kappa\tilde c_0}{2}c+\left\llbracket\tilde\sigma\p{1+r}\p{1+\ln\phi}\right\rrbracket\,.
\end{eqnarray}
We thus obtain
\begin{eqnarray}
\delta W&=&\int\!d^2 u\,\p{\sqrt{a}\,I_1\delta c+I_2\delta\sqrt{a}}\,,
\end{eqnarray}
where we have defined
\begin{eqnarray}
I_2&=&\sigma_0+\sigma_1\psi+\frac{\sigma_2}{2}\psi^2-\frac{\kappa}{2}\p{c_0+\tilde c_0\psi}c\nonumber\\
&+&\frac{\kappa}{4}c^2+\frac{k}{2}\p{ec-r}\p{r+ec+2}\,.
\end{eqnarray}
Note that the contribution of the term between double square brackets has vanished, as in our calculation of $\mathbf{\Sigma}$.

Thanks to the relations
\begin{eqnarray}
\delta\sqrt{a}&=&\sqrt{a}\,\,\bm{t}^\alpha\cdot\delta\bm{t}_\alpha\,,\\
\delta c&=&a^{\alpha\beta}\p{\partial_\alpha\bm{n}}\cdot\delta\bm{t}_\beta-\bm{t}^\alpha\cdot\delta\p{\partial_\alpha\bm{n}}\,,\\
\delta\bm{n}&=&-\p{\bm{n}\cdot\delta\bm{t}_\beta}\bm{t}^\beta\,,
\end{eqnarray}
$\delta W$ can be expressed only in terms of $\delta\bm{t}_\alpha$. Performing two integrations by parts and using the relations \cite{Aris}
\begin{eqnarray}
\bm{t}_{\alpha|\beta}&=&b_{\alpha\beta}\bm{n} \label{dct}\,,\\
\bm{n}_{|\alpha}&=&\partial_\alpha\bm{n}=-b_{\alpha\beta}\bm{t}^\beta\label{dcn}\,,
\end{eqnarray}
where $g_{|\alpha}$ denotes the covariant derivative (associated with the metric $a_{\alpha\beta}$) with respect to $u^\alpha$ of a function $g$ defined on the surface \cite{Aris}, 
we obtain
\begin{eqnarray}
\delta W=-\!\int\!dA\!&\Big[&\!\p{a^{\alpha\beta}I_1 - b^{\alpha\beta}I_2 }\bm{t}_\alpha\nonumber\\
&+&a^{\alpha\beta}\p{\partial_\alpha I_2}\bm{n}\Big]_{|\beta}\cdot\delta\bm{R}\,.
\label{deltaw}
\end{eqnarray}
Identifying (\ref{deltaw}) with (\ref{travail}) for any infinitesimal virtual deformation $\delta \bm{R}$, we obtain the sought force density:
\begin{equation}
\bm{q}=\psq{\p{a^{\alpha\beta}I_1 - b^{\alpha\beta}I_2 }\bm{t}_\alpha+a^{\alpha\beta}\p{\partial_\alpha I_2}\bm{n}}_{|\beta}\,.
\label{idnq}
\end{equation}
Performing the covariant derivative with respect to $u^\beta$ in (\ref{idnq}), using (\ref{dct})--(\ref{dcn}) and the relations \cite{Aris}
\begin{eqnarray}
b^\alpha_{\beta|\alpha}&=&\partial_\beta c\,,\\
b^{\alpha\beta}b_{\beta\alpha}&=&c^2-c_1 c_2\,,
\end{eqnarray}
and taking the scalar product of $\bm{q}$ with $\bm{t}_\alpha$ (resp. $\bm{n}$) finally leads to (\ref{q_tg}) (resp. (\ref{q_n})).


\begin{thebibliography}{11}

\bibitem{Alberts_book}
B. Alberts, A. Johnson, J.  Lewis, M. Raff, K.
Roberts, P. Walter, \textit{Molecular Biology of the Cell}
(Garland, New York, 2002), 4th ed.

\bibitem{Mouritsen_book}
O. G. Mouritsen, \textit{Life---as a matter of
fat} (The frontiers collection, Springer, Berlin, 2005). 

\bibitem{Helfrich73}
W. Helfrich, Z. Naturforsch. C {\bf 28}, 693 (1973).

\bibitem{Helfrich84}
W. Helfrich, R.-M. Servuss, Nuovo Cimento D \textbf{3}, 137 (1984).

\bibitem{Fournier01}
J.-B. Fournier, A. Ajdari and L. Peliti, Phys. Rev. Lett. \textbf{86}, 4970 (2001).

\bibitem{Helfrich78}
W. Helfrich, Z. Naturforsch. A {\bf 33}, 305 (1978).

\bibitem{Brochard75}
F. Brochard and J.-F. Lennon, J. Physique {\bf 36}, 1035 (1975).

\bibitem{Engelhardt85}
H. Engelhardt, H. P. Duwe and E. Sackmann,
J. Physique Lett. \textbf{46}, 395 (1985).

\bibitem{Meleard92}
P. M\'el\'eard, J. F. Faucon, M.  D. Mitov, and P.
Bothorel, Europhys. Lett. {\bf19}, 267 (1992).

\bibitem{Lim02}
Lim G. H. W., Wortis M. and Mukhopadhyay R., PNAS \textbf{99}, 16766 (2002). 

\bibitem{Derenyi02}
I. Der\'enyi, F. J\"ulicher, J. Prost, Phys. Rev. Lett. \textbf{88}
238101 (2002).

\bibitem{Baumgart03}
T. Baumgart, S. T. Hess and W. W. Webb, Nature \textbf{425}, 821 (2003).

\bibitem{Evans90}
E. Evans and W. Rawicz, Phys. Rev. Lett. \textbf{64}, 2094 (1990).

\bibitem{Evans80}
E. Evans, Biophys. J. \textbf{30}, 265 (1980).

\bibitem{Svetina89}
S. Svetina and B. {\v Z}ek{\v s}, Eur. Biophys. J. \textbf{17}, 101 (1989).

\bibitem{Wiese92}
W. Wiese, W. Harbich and W. Helfrich, J. Phys.: Condens. Matter \textbf{4}, 1647 (1992). 

\bibitem{Svetina85}
S. Svetina, M. Brumen and B. {\v Z}ek{\v s}, Stud. Biophys.
\textbf{110}, 177 (1985).

\bibitem{Miao94}
L. Miao, U. Seifert, M. Wortis, H.-G. D{\"o}bereiner, Phys. Rev. E
\textbf{49}, 5389 (1994).

\bibitem{Seifert_book}
U. Seifert, L. Miao, H.-G. D\"obereiner, and M. Wortis, in \textit{The Structure 
and Conformation of Amphiphilic Membranes}, Vol. 66 of Springer Proceedings in Physics, edited by R. Lipowsky, D. Richter, and K. Kremer 
(Springer, Berlin, 1991), pp. 93-96. 

\bibitem{Evans74}
E.A. Evans, Biophys. J. \textbf{14}, 923 (1974).

\bibitem{Khalifat08}
N. Khalifat, N. Puff, S. Bonneau, J.-B. Fournier and M. I. Angelova, Biophys. J. \textbf{95}, 4924 (2008).

\bibitem{Fournier09} J.-B. Fournier, N. Khalifat, N. Puff and M. I. Angelova, 
Phys. Rev. Lett. \textbf{102}, 018102 (2009).

\bibitem{Mabrouk09}
E. Mabrouk, D. Cuvelier, F. Brochard-Wyart, P. Nassoy and M. H. Li, PNAS \textbf{106}, 7294 (2009).

\bibitem{Evans94}
E. Evans, A. Yeung, Chem. Phys. Lipids \textbf{73}, 39 (1994).

\bibitem{Seifert93}
U. Seifert and S. A. Langer, Europhys. Lett. \textbf{23}, 71 (1993).

\bibitem{Kabaso10}
D. Kabaso, R. Shlomovitz, T. Auth, V. L. Lew and N. S. Gov, Biophys. J \textbf{99} 808 (2010).

\bibitem{Napoli10}
G. Napoli and L. Vergori, J. Phys. A: Math. Theor. \textbf{43} 445207 (2010).

\bibitem{FuturMig} A.-F. Bitbol, J.-B. Fournier, M. I. Angelova and N. Puff, 
Accepted for publication in J. Phys.: Condens. Matter (to be published in May 2011).

\bibitem{Muller07}
M. M. M\"uller and M. Deserno, Phys. Rev. E \textbf{76} 011921 (2007).

\bibitem{Bitbol10}
A.-F. Bitbol, P. G. Dommersnes and J.-B. Fournier, Phys. Rev. E \textbf{81} 050903(R) (2010).

\bibitem{Capovilla02}
R. Capovilla and J. Guven, J. Phys. A \textbf{35}, 6233 (2002).

\bibitem{Fournier07}
J.-B. Fournier, Soft Matter \textbf{3}, 883 (2007).

\bibitem{Seifert95}
U. Seifert, Z. Phys. B \textbf{97}, 299 (1995).

\bibitem{Cai95}W. Cai and T. C. Lubensky, Phys. Rev. E \textbf{52}, 4251 (1995).

\bibitem{Miao02}L. Miao, M. A. Lomholt and J. Kleis, Eur. Phys. J. E \textbf{9}, 143 (2002).

\bibitem{Doi_book}
M. Doi, \textit{Introduction to polymer physics} (Oxford Science Publications, 1995).

\bibitem{Safran_book}
S. A. Safran, {\em Statistical Thermodynamics of Surfaces,
Interfaces and Membranes\/} (Addison-Wesley, Reading, MA, 1994).

\bibitem{Petrov84}A. G. Petrov and I. Bivas, Prog. Surf. Sci. \textbf{16}, 389 (1984).

\bibitem{Shkulipa06}S. A. Shkulipa, W. K. den Otter and W. J. Briels, J. Chem. Phys. \textbf{125}, 234905 (2006).

\bibitem{Aris}R. Aris, \textit{Vectors, tensors, and the basic equations of fluid dynamics} (Dover, New York, 1989).

\bibitem{Lomholt05}M. A. Lomholt, P. L. Hansen and L. Miao, Eur. Phys. J. E \textbf{16}, 439 (2005).

\bibitem{Jenkins77}J. T. Jenkins, J. Math. Biol. \textbf{4}, 149 (1977).

\bibitem{Powers10} T. R. Powers, Rev. Mod. Phys. \textbf{82}, 1607 (2010).

\bibitem{Capovilla02b} R. Capovilla, J. Guven and J. A. Santiago, Phys. Rev. E \textbf{66}, 021607 (2002).   

\bibitem{Capovilla02c}R. Capovilla, J. Guven, Phys. Rev. E \textbf{66}, 041604 (2002).   

\bibitem{Deserno07}M. Deserno, M. M. M\"uller and J. Guven, Phys. Rev. E \textbf{76}, 011605 (͑2007͒).

\bibitem{Fournier08} J.-B. Fournier and C. Barbetta, Phys. Rev. Lett. \textbf{100}, 078103 (2008).

\bibitem{Barbetta10} C. Barbetta, A. Imparato and J.-B. Fournier, Eur. Phys. J. E \textbf{31}, 333 (2010).

\bibitem{Barbetta09} C. Barbetta and J.-B. Fournier, Eur. Phys. J. E \textbf{29}, 183 (2009).

\bibitem{Muller05} M. M. M\"uller, M. Deserno and J. Guven, Phys. Rev. E \textbf{72}, 061407 (2005).

\bibitem{Muller05b} M. M. M\"uller, M. Deserno and J. Guven, Europhys. Lett. \textbf{69}, 482 (2005). 

\end{thebibliography}
\end{document}